\documentclass[12pt]{iopart}

\usepackage{graphicx}
\usepackage{dcolumn}
\usepackage{bm}

\usepackage{gensymb}
\usepackage[normalem]{ulem}
\usepackage[utf8]{inputenc}
\usepackage[T1]{fontenc}
\usepackage{mathptmx}
\usepackage{etoolbox}
\usepackage{xcolor}
\makeatletter
\def\@email#1#2{%
 \endgroup
 \patchcmd{\titleblock@produce}
  {\frontmatter@RRAPformat}
  {\frontmatter@RRAPformat{\produce@RRAP{*#1\href{mailto:#2}{#2}}}\frontmatter@RRAPformat}
  {}{}
}%
\makeatother

\begin{document}

\title{Low-vibration cryogenic test facility for next generation of ground-based gravitational-wave observatories}

\author{D. P. Kapasi, T. G. McRae, J. Eichholz, P. A. Altin, D. E. McClelland and B. J. J. Slagmolen}

\address{OzGrav-ANU, ARC Centre for Gravitational Astrophysics, College of Science, The Australian National University, ACT 2601, Australia. }

\ead{disha.kapasi@anu.edu.au}
\vspace{10pt}
\begin{indented}
\item[]August 2024
\end{indented}

\begin{abstract}
We present the design and commissioning of a cryogenic low-vibration test facility that measures displacement noise from a gram-scale silicon cantilever at the level of 10$^{-16}\, \mathrm{m/\sqrt{Hz}}$ at 1\,kHz. A volume of $\sim$36 litres is enclosed by radiation shields cooling an optical test cavity that is suspended from a multi-stage pendulum chain providing isolation from acoustic and environmental noise. This 3\,kg test cavity housing a crystalline silicon cantilever is radiatively cooled to 123\,K in 41 hours and held at that temperature over many months with a relative temperature stability of $\pm$1\,mK. The facility is capable of interferometrically measuring temperature-dependent broadband displacement noise between 50\,Hz and 10\,kHz where current and future ground-based gravitational wave observatories are most sensitive. With suitable cantilever design, the cryogenic facility we describe here will allow for the measurement of broadband thermal noise in crystalline silicon at 123\,K. This will guide the design of suspensions in planned future cryogenic ground-based gravitational-wave detectors such as LIGO Voyager and the Einstein Telescope. This facility is also suitable for the testing of new mirror coatings at cryogenic temperatures.
\end{abstract}

\vspace{2pc}
\noindent{\it Keywords\/}: gravitational waves, cryogenics, silicon, thermal noise, future detectors

\submitto{\CQG}
\maketitle
\section{Introduction}

Ground-based gravitational-wave observatories are large complex opto-mechanical systems measuring astrophysical signals from binary compact mergers since 2015 \cite{gw150914,O1-O2catalogue, lvccollab2, O3full, Multimessenger2017, GW170817}. The thermal environment of the test masses sets a physical limit on their sensitivity. The power spectral density of thermally-induced Brownian motion in an object is proportional to its temperature and mechanical loss \cite{Saulson1990, Callen:1952}. Present ground-based observatories like advanced LIGO and advanced Virgo have test mass optics and suspensions made of fused silica that suffer from large substrate Brownian noise at $\sim$40\,K and poor thermal conductivity when cooled to cryogenic temperatures \cite{Schroeter:2007}. Thus, to reduce thermal noise impacts, future designed terrestrial gravitational-wave observatories such as LIGO Voyager and the Einstein Telescope (ET) are planned to operate at cryogenic temperatures with crystalline silicon test masses and suspensions \cite{Adhikari:2020, ET1, ET2}. The Neutron-star Extreme Matter Observatory (NEMO) is a design study for a high-frequency gravitational-wave detector aimed at measuring the post-merger ringdowns from binary coalescences using cryogenically cooled crystalline silicon as test masses \cite{NEMO}.

The mechanical loss ($\phi$) in silicon is lower than fused silica at cryogenic temperatures, thereby reducing Brownian noise in the substrates. Ringdown measurements of silicon cantilevers with different surface treatments (wet and dry etching) have shown this loss to be as low as $3\times10^{-8}$ at 10\,K \cite{Nawrodt:2013uca, Reid:2004yqx}. On the contrary, fused silica has a total loss to be $10^{-3}$ at 30\,K \cite{Schroeter:2007}. Silicon's coefficient of linear thermal expansion (CTE) crosses zero at 123\,K and 18\,K, naturally making these temperatures good choices for operating points, as thermo-elastic noise and thermal distortions on the mirror surface are strongly suppressed by the absence of thermal expansion \cite{shiela:2003, siliconthermalnoise, Degallaix_2014}. As silicon becomes colder, its thermal conductivity gets higher and reaches a peak at around 20\,K \cite{Silicon_conductivity}. Even at room temperature, silicon has thermal conductivity $\sim$2 orders of magnitude higher than fused silica, and thereby reduces temperature gradients generated by laser beam absorption. This allows heat to be efficiently extracted from the centre of the test mass which is critical in high power operation of the detectors as it minimises thermal distortions. 

\begin{figure}
\centering
\includegraphics[width=0.7\linewidth]{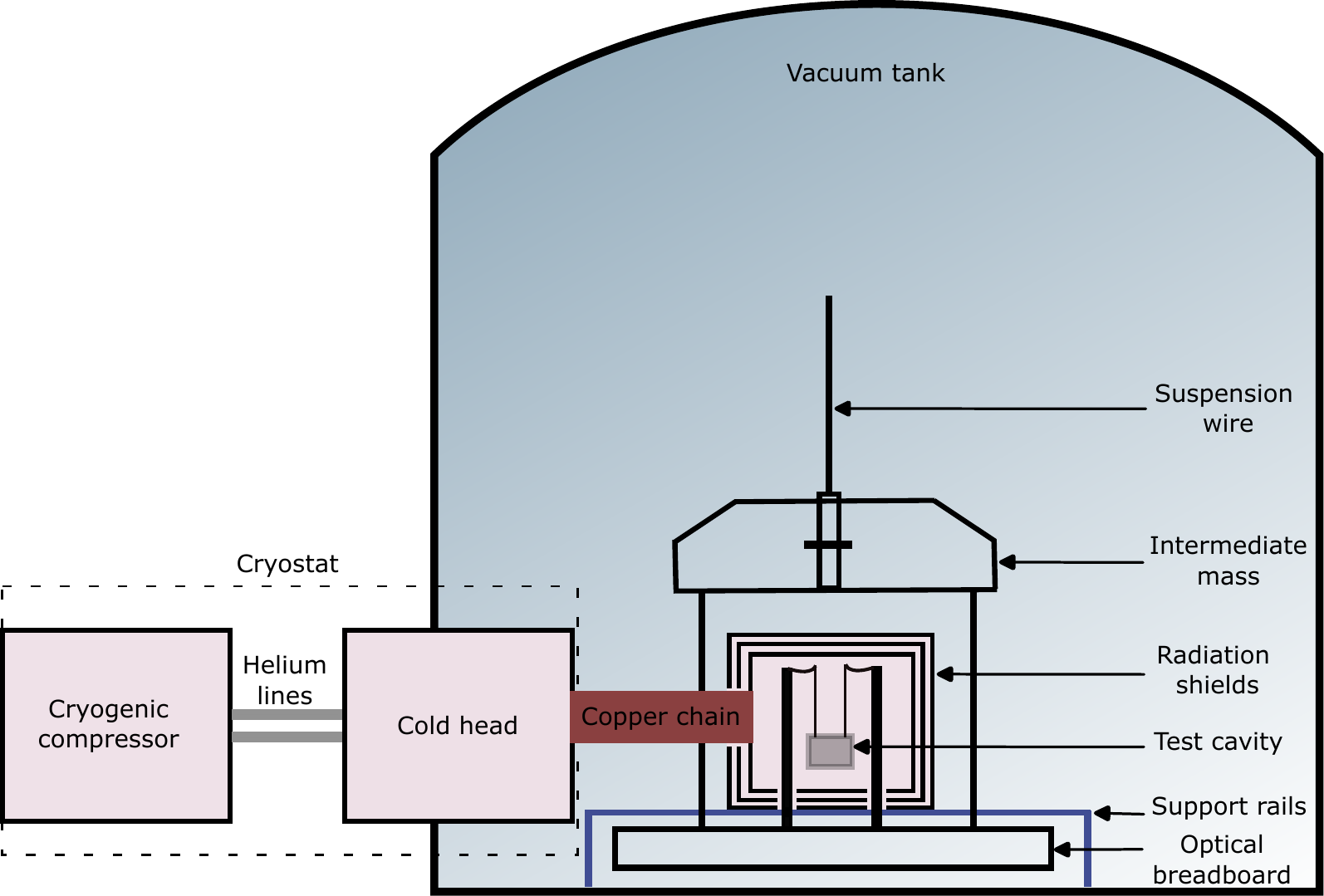}
\caption{Overview of the cryogenic facility. A large optical breadboard is suspended from the intermediate mass via four wires. The test cavity is suspended from this breadboard, and is radiatively cooled by multi-layer thermal shields. The cryogenic infrastructure sits on support rails, therefore making it independent of the pendulum (i.e. seismic isolation chain). A detailed view of the infrastructure inside the vacuum envelope is shown in Figure \ref{fig:breadboard}.}
\label{fig:blockfig}
\end{figure}

\begin{figure}
\centering
\includegraphics[width=0.8\linewidth]{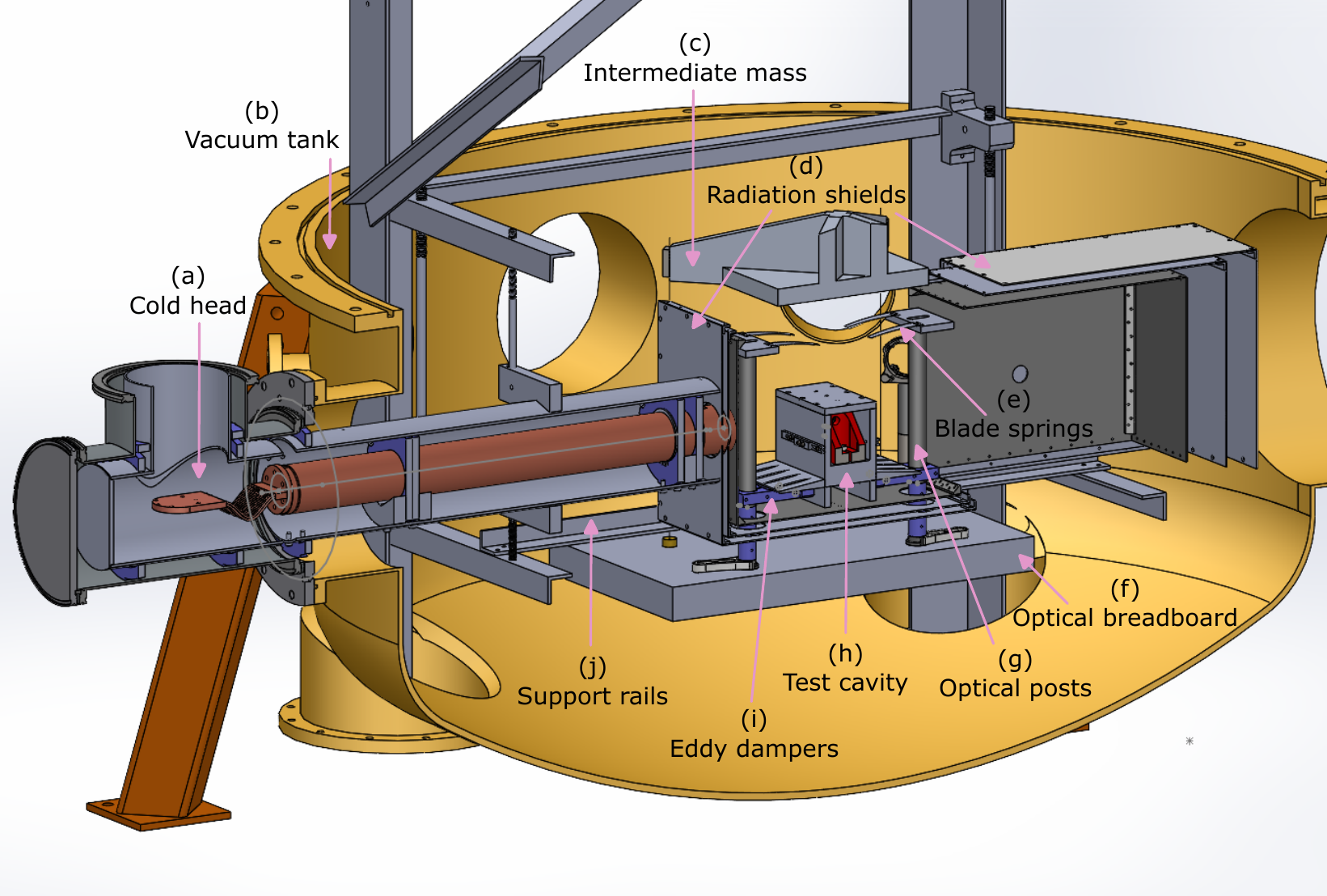}
\caption{CAD drawing showing the overview of the cryogenic infrastructure. (a) The coldhead attaches to the copper heat pipe assembly via a flexible copper braid (see Figure \ref{fig:coldhead} for details). (b) Vacuum tank enclosing both the seismic isolation chain and the cryogenic infrastructure. (c) Intermediate mass in the seismic isolation chain. (d) Thermal radiation shields for radiative transfer of heat from the test cavity. In this cross section view, the shields are retracted. (e) Blade springs suspending the optical test cavity. (f) Optical breadboard suspended by four wires from the intermediate mass that remains seismically isolated. (g) Optical posts with a thermal break. (h) Optical test cavity suspended from maraging blade springs attached to the breadboard via optical posts (see Figure \ref{fig:testcavity} for assembly of the test cavity). (i) Magnets for eddy current damping mounted in PEEK supports to damp cavity motion relative to the optical breadboard. (j) Shield support rail attached to frame.}
\label{fig:breadboard}
\end{figure}

This paper presents the design and construction of a low-vibration noise cryogenic infrastructure to cool an optical test cavity (TC) incorporating a silicon cantilever to 123\,K where the CTE for silicon crosses zero. Section \ref{section:cryo} provides details of the cryogenic infrastructure. In particular, we discuss the modelling of the facility using a simple model (section \ref{section:thermalmodel}) to estimate the cooldown time for the TC to reach 123\,K and in the subsequent subsections we provide details of the various subsystems. Section \ref{section:results} discusses the performance of the cryogenic facility and presents the first displacement noise results from the TC in subsections \ref{section:performance} and \ref{section:displacement}, respectively. 

\section{Cryogenic infrastructure}
\label{section:cryo}

\subsection{Overview}
Figure \ref{fig:blockfig} shows a simplified overview of the cryogenic infrastructure that was constructed in this work. A 40\,W cryostat is used to provide cooling; it consists of a compressor and a coldhead unit. The compressor is connected to the coldhead via flexible high and low pressure helium lines bringing the coldhead to cryogenic temperatures. The heat is transported from the radiation shields housed inside the vacuum tank by means of a copper heat pipe. The multi-layer shields enclose the TC that is cooled to 123\,K by radiative heat flow, thereby providing a low-vibration environment for sensitive opto-mechanical experiments. The facility  is used to directly measure broadband displacement noise in crystalline silicon at levels of $\sim$$10^{-16}$ $\mathrm{m}/\sqrt{\mathrm{Hz}}$ at cryogenic temperatures.

Room temperature off-resonance thermal noise measurements from silicon cantilevers have been described in detail in previous works \cite{ThanhKapasi:2023,NguyenThesis2015,Nguyen:2015}. The measured displacement noise from a 100$\mu$m silicon flexure that is part of an optical TC was recorded at the level of $10^{-15}$ $\mathrm{m}/\sqrt{\mathrm{Hz}}$ at 100\,Hz with a total loss of $1.8\times10^{-5}$ at the fundamental resonance \cite{ThanhKapasi:2023}. For the results presented in section \ref{section:displacement}, a thermal noise model was built using the Fluctuation Dissipation Theorem to predict displacements in gram-scale silicon cantilevers \cite{Callen:1952}. The model consists of frequency-independent mechanical loss and frequency-dependent loss, which is assumed to be dominated by thermo-elastic damping. Other noises from extrinsic sources and experimental readout schemes are also considered to predict the total noise \cite{KapasiThesis2024}. 

With the need for a low-noise cryogenic infrastructure, cooling the TC via conduction was eliminated as it posed a high risk of introducing mechanical noise from the cryostat, and short circuiting the suspension system. This led to incorporating radiative heat transfer for cooling the TC using multi-layer thermal shields as it is a contactless mode of heat transport. Figure \ref{fig:breadboard} shows the technical drawing of the cryogenic infrastructure integrated with the seismic isolation chain \cite{BramThesis2004}. The various subsystems labelled in Figure \ref{fig:breadboard} will be discussed in subsequent sections. In particular, there are detailed descriptions of the copper heat pipe integration, the radiation shields and the TC in sections \ref{section:copperchain}, \ref{section:shields} and \ref{section:testcavity} respectively.\\ 

\subsection{Cooldown model}
\label{section:thermalmodel}
A simple model is constructed to estimate the cooldown time for the TC in the cryogenic infrastructure. The model consists of a copper heat pipe utilising conductive heat flow and three layers of thermal shields made from aluminium for radiatively cooling the TC to 123\,K. The two outer shields are polished resulting in a low emissivity to minimise heat transfer from the environment to the TC. The inner shield is hard anodised to increase the emissivity to radiatively transfer heat away from the TC\footnote{A conservative value of $\epsilon_{low} = 0.2$ and $\epsilon_{high} = 0.5$ is used in the cooldown model.}.

One end of the copper heat pipe is attached to the coldhead of the cryostat via a flexible copper braid. The entire heat pipe is $\sim$850\,mm long with a diameter of 70\,mm and weighs approximately 28\,kg in total. The dominant mode of heat transfer through this heat pipe is via conduction. The rate of conductive heat flow through a uniform solid with cross-sectional area $A$, length $l$, and thermal conductivity $\kappa$ is given by Fourier's law \cite{Holman}
\begin{equation}
P_{cond} = \frac{\kappa\,A}{l}(T_{2} - T_{1}),
\label{eqn:conductivity}
\end{equation}
where $T_{1}$ and $T_{2}$ are the temperatures of the solid at the ends with $T_2 > T_1$. 

The other end of the copper heat pipe is connected to the innermost radiation shield, modelled as a rectangular box where all faces of the inner shield are cooled via conduction. The inner shield in turn cools the cavity to cryogenic temperatures via radiation. The heat emitted by an object with emissivity $\epsilon$ and surface area $A$ is governed by the Stefan-Boltzmann law \cite{Holman}
\begin{equation}
P_{rad} = {\sigma\,\epsilon\,A}\,T^{4},
\label{eqn:radiation}
\end{equation}
where $\sigma$ is the Stefan-Boltzmann constant 
and $T$ is the surface temperature of the body. The consecutive layers of the radiation shield are separated by polyether ether ketone (PEEK) spacers that reduces heat flow via conduction between the shields \cite{PEEK}. 

\begin{figure}
\centering
\includegraphics[width=0.5\linewidth]{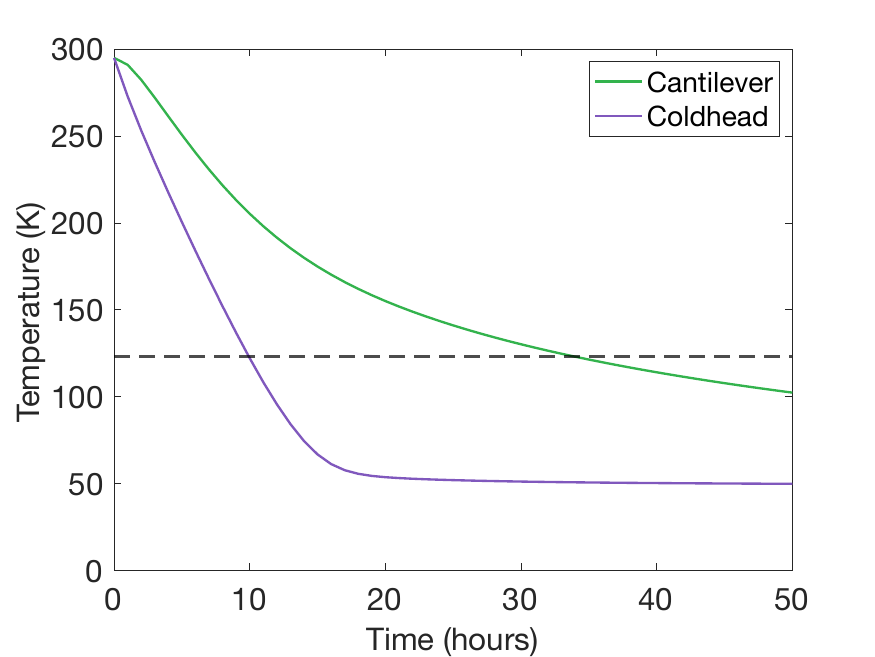}
\caption{Thermal modelling of the cryogenic infrastructure. The green and purple lines show the estimated rate of cooldown for the cantilever and the coldhead respectively. The dashed line indicates 123\,K.}
\label{fig:thermalmodel}
\end{figure} 

When calculating the heat load on the inner radiation shield, geometric shape factors between the different surfaces are considered. Using the network analysis method for radiative heat transfer and conductive heat flow, a simple model is built to estimate cooldown times for the coldhead and the cantilever \cite{Oppenheim,Holman}. First, the surface area, volume and mass of each component of the system is determined. These parameters are used to construct heat flow equations for the model above using the temperature-dependent heat capacity of the relevant materials. These ordinary differential equations are then solved using the backward differentiation method with the initial state of the system set to 295\,K at $t$\,=\,0. The temperature of the coldhead and the cantilever are evaluated at intervals of 60\,s. For a realistic prediction, the thermal model also takes into account the cooling power of our cryostat given in the manufacturer's specifications between 30\,K and 80\,K. Beyond 80\,K the cooling power is extrapolated using a two-term exponential model that is fitted to the data provided by the manufacturer. At steady state, the heat load on the TC is evaluated to be $\sim$3\,W at 123\,K. Figure \ref{fig:thermalmodel} shows the estimated cooldown curves for the coldhead and the cantilever. The model predicts that the cantilever will take around 35 hours to reach the target temperature of 123\,K (indicated by the dashed line in the figure).

\subsection{Cryostat} 
\label{section:cryostat}

A pulse-tube cryocooler from CryoMech is used for the cryostat. 
Since there are no moving parts at the cold end of a pulse-tube cryocooler, it is an ideal choice for a low-vibration cryogenic environment. The compressor (CP103A) is air-cooled and is housed in a plant room adjacent to the main lab. 
The CryoMech PT-60RM coldhead has a cooling capacity of $\sim$6\,W at 40\,K. The remote motor for the coldhead is mounted on the wall near the coldhead ($\sim$1\,m away) in the main lab and is powered via the cryogenic compressor \cite{Cryomech}. 

\subsection{Heat pipe}
\label{section:copperchain}

\begin{figure}
\centering
\includegraphics[width=0.6\linewidth]{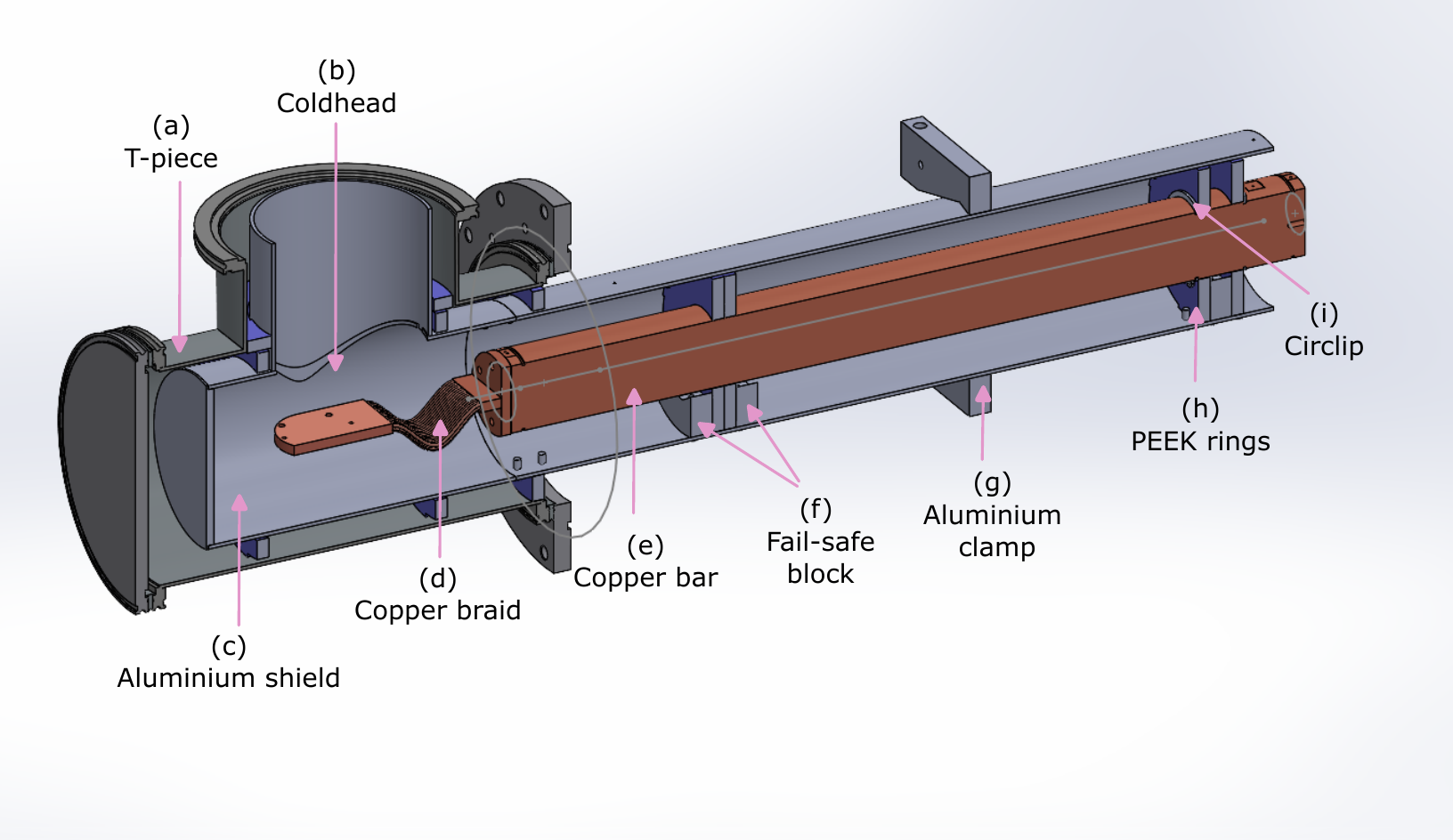}
\caption{Cross-section of the copper heat pipe assembly. (a) 8\,inch ISO T-piece vacuum flange assembly for cryostat. (b) Coldhead attachment point to the braid. (c) Aluminium shield to reflect thermal radiation at 300\,K. (d) OFHC flexible copper braid to reduce mechanical coupling and allow thermal expansion of the heat pipe. (e) Copper bar to transport heat to/from the test cavity. (f) Fail-safe blocks rest several millimetres below the copper bar to catch it if the PEEK were to fail with repeated thermal cycling. (g) The aluminium clamp fixes the aluminium shield to the vacuum infrastructure. (h) PEEK rings to support the weight of the copper bar and reduce heat flow via thermal conductivity between aluminium shield and the copper bar. (i) Stainless steel circlips to minimise the movement of the copper bar due to thermal expansion at the end closer to the experiment.}
\label{fig:coldhead}
\end{figure}

Figure \ref{fig:coldhead} shows the copper heat pipe assembly. The cyrostat housing the coldhead is 70.6\,mm in diameter and mates to a copper spacer. This copper spacer is 80\,mm in length, and weighs around 2.8\,kg. At cryogenic temperatures, a dominant source of heat loss is poor conduction between joints/interfaces. To enhance the surface contact between different parts of the heat pipe, a thin layer of indium foil (100$\mu$m, 99.999$\%$ purity) is compressed between the joints providing a high thermal conductivity interface. These indium sheets are  first laser cut to match the dimensions and bolt pattern of the respective joints, and then wiped with high purity isopropanol before installation. To stop similar materials bolted together becoming cold welded, a small amount of cryogenic grease (Apiezon-N) was used \cite{Apeizon}.

The copper spacer is then connected to a copper braid made from oxygen free high thermal conductivity (OFHC) copper. The braid consists of 48\, braids where each of these individual braid is 1\,mm in diameter and 100\,mm in length. By using a flexible copper braid one can reduce the transfer of any residual mechanical vibrations from the cryostat to the experiment. The other end of the copper braid connects to a copper bar (Cu bar) via six vented M6 stainless steel bolts and washers as shown in Figure \ref{fig:coldhead}. The copper bar is 735\,mm long and weighs 25\,kg. This makes it a large thermal mass for the cryogenic infrastructure which can keep the experiment at 123\,K for 11\, hours after the cryostat is turned off (see Figure \ref{fig:cooldown2021}). The copper bar is fixed at the inner shield at the far end and this forces the expansion and contraction of the copper bar to occur closer to the braid. The copper bar is supported by three PEEK rings of 15\,mm thickness. The PEEK rings make a three-point contact with the aluminium shield housing the copper bar. This reduces the conductive heat flow from the aluminium shield to the copper bar. Stainless steel circlips (retaining rings) are installed on the copper bar adjacent to the outer side of the PEEK rings to constrain the movement of the copper bar relative to the PEEK. 
Aluminium blocks cut with a recess to catch the copper bar are placed between the circlips in case of catastrophic failure of the PEEK with repeated thermal cycling. These fail-safe blocks (item (f) in Figure \ref{fig:coldhead}) are bolted firmly to the aluminium shield and do not physically touch the copper bar during usual operation. 
The aluminium shield containing the heavy copper bar is clamped to the frame (item (g) in Figure \ref{fig:coldhead}) that supports the pendulum, i.e., the seismic isolation chain. This restricts the movement of the aluminium shield with respect to the copper bar. To reflect thermal energy at 300\,K, the shield is wrapped in two layers of Mylar (25.4$\mu$m thickness, 99.997$\%$ reflective). Prior to installation, these sheets were baked out at 80\,$\degree$C to remove moisture. 


\subsection{Radiation shields}
\label{section:shields}

To keep the cryogenic infrastructure mechanically isolated from the pendulum, two aluminium rails (item (j) in Figure \ref{fig:breadboard}) are installed over the large breadboard, with a 10\,mm clearance from the optical breadboard on which the radiation shields rest. There are three radiation shields (outer, intermediate, and inner) sitting concentrically on the aluminium rails and separated by PEEK spacers at the four corners on the base. The spacers are also designed not to be a tight fit and therefore allow strain relief between the shields due to thermal contraction as the system is cooled down. 

Each shield is made from 1.8\,mm thick 5000 series aluminium suitable for anodisation. The gap between consecutive shield layers is 10\,mm, with a 1 inch hole on the front and back surfaces for routing the optical beam to the experiment. The edges of the box are bolted together using stainless steel spring washers and internal struts running the length of the joint to enhance the thermal contact for better heat flow between adjacent faces of the box.

For the inner shield, the main aim is to maximise heat flow to the TC through radiative heat transfer. Two ways to increase the rate of radiative heat flow are by choosing a material with high emissivity and increasing the surface area (see equation \ref{eqn:radiation}). The inner shield is hard anodised with a depth of 50\,$\mu$m that increases the surface area for radiative cooling, and further increases the emissivity. The use of dye (organic or inorganic) to increase the emissivity was eliminated as it is a potential source of contamination of optics in the vacuum due to outgassing. While assembling the inner shield, indium foil is used at each of the angled joints to ensure good surface contact and therefore increase the heat flow between the adjacent surfaces. 

\subsection{Test cavity}
\label{section:testcavity}

\begin{figure}
\centering
\includegraphics[width=0.6\linewidth]{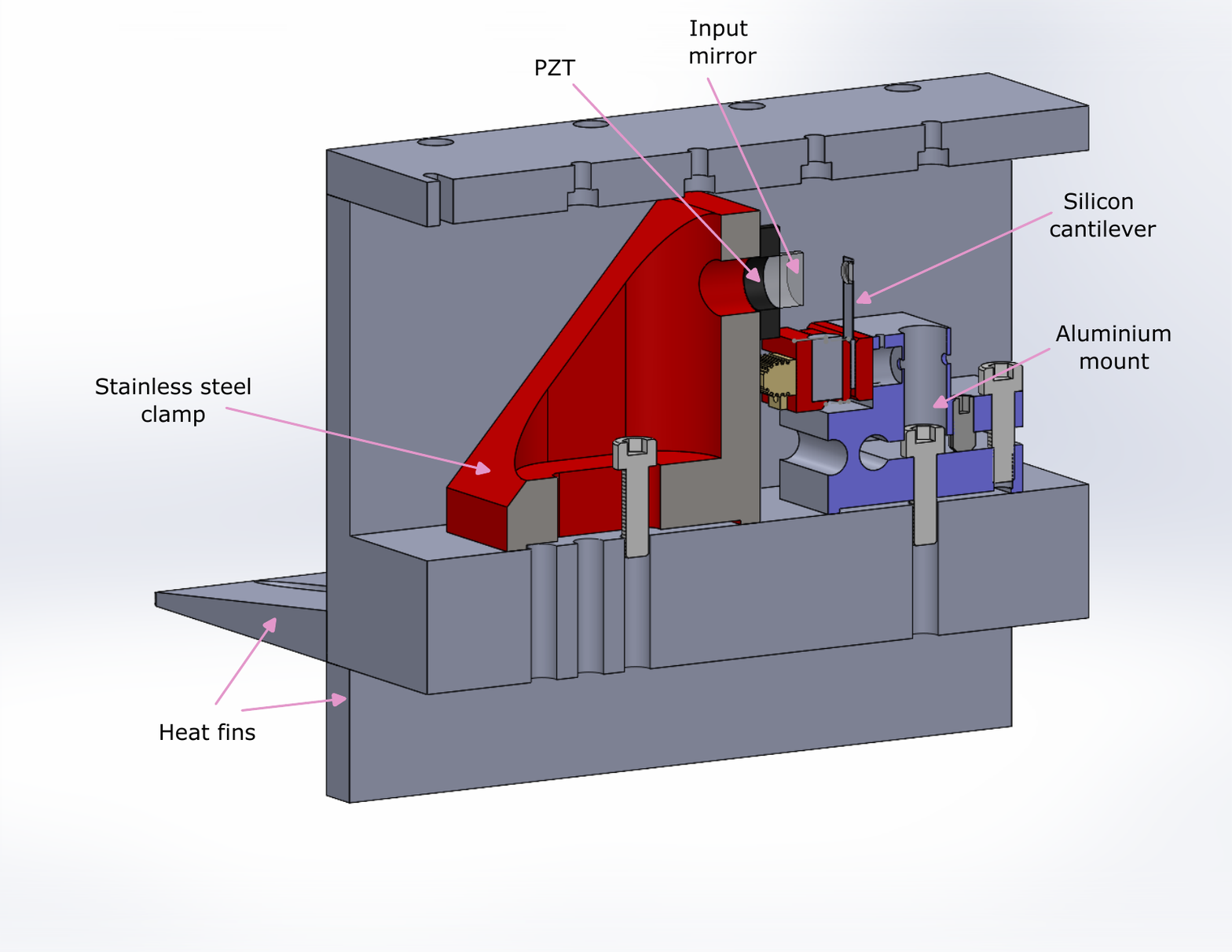}
\caption{Cut away showing the assembly of the test cavity. A monolithic U-shaped aluminium base houses a plano-concave cavity formed between the input mirror on a PZT stack and an output mirror glued to a silicon cantilever.} 
\label{fig:testcavity}
\end{figure}

The suspended TC that is cooled to cryogenic temperatures is a plano-concave Fabry-P\'erot cavity with a nominal length of 10\,mm and finesse 1000. The 1/2 inch input mirror is glued to a piezoelectric stack (PZT) using a cryogenic adhesive (Stycast 2850FT) and the 1/4 inch output flat mirror is glued to a 500\,$\mu$m dual side polished silicon cantilever with a clearance hole at the free end that allows beam transmission for cavity diagnostics as shown in Figure \ref{fig:testcavity} \cite{Stycast, Stycast2}. The cantilever used for the study here is a prototype aimed at testing the capabilities of the cryogenic facility. Our cantilever is fabricated closely to represent the geometry for the ribbon suspension for LIGO Voyager which is designed with a nominal width of 10\, mm and a thickness of 500\,$\mu$m \cite{Adhikari:2020}. The prototype silicon cantilever was fabricated from a $\langle 1 0 0 \rangle $ silicon wafer and has a fundamental resonant frequency (including mirror) of 1450\,Hz for the bending mode which was verified using finite element analysis. The two mirrors of the TC are mounted on a common base plate with heat fins that is suspended from a multi-stage passive pendulum to attenuate environmental noise due to ground motion and other anthropogenic activities coupling into the experiment \cite{BramThesis2004,ConorThesis2011}. 

\begin{figure}
\centering
\includegraphics[width=0.6\linewidth]{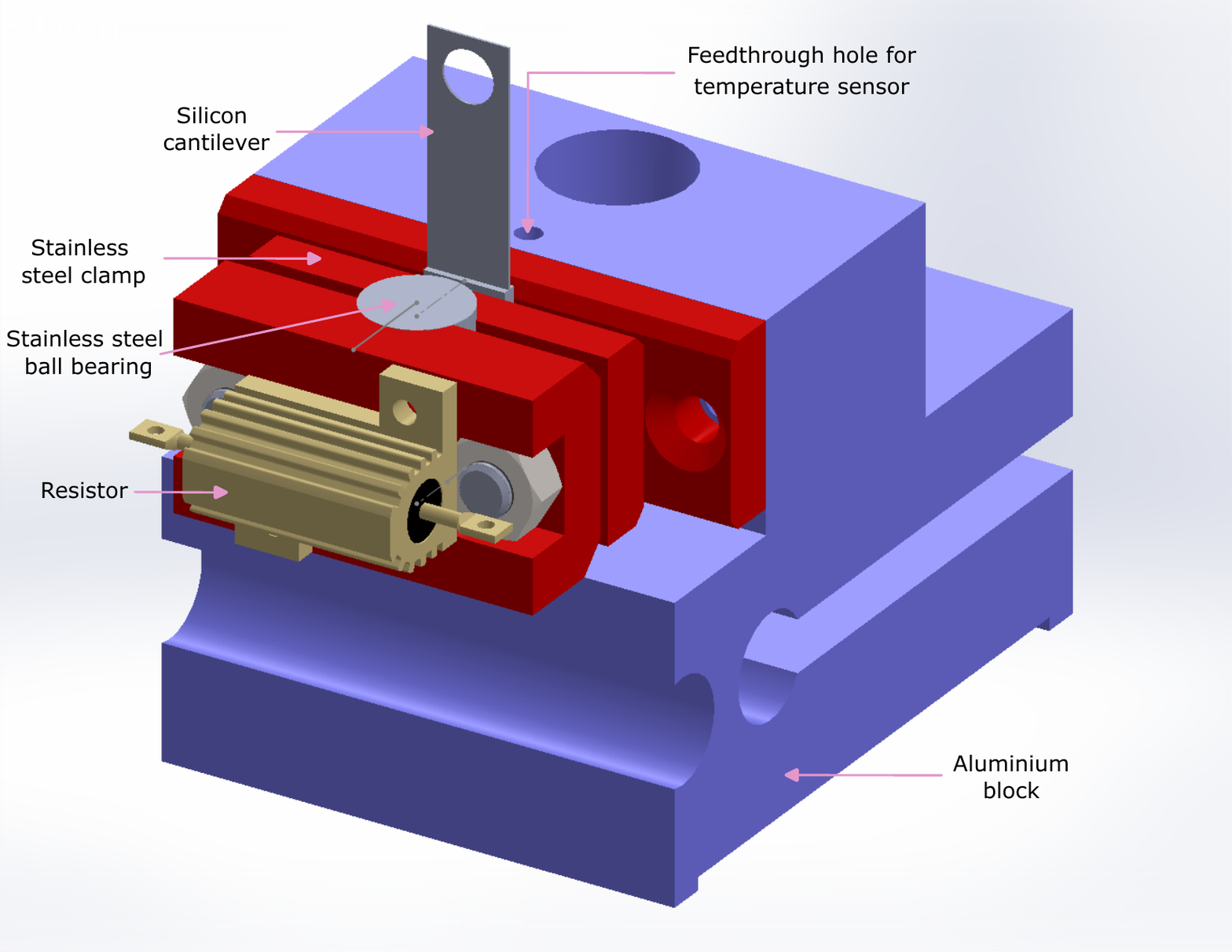}
\caption{Clamp design for mounting of the silicon cantilever. The cantilever is clamped between stainless steel plates and a cylindrical bearing assists in applying even pressure to the plates. The temperature sensor is glued to the inner stainless steel plate. }
\label{fig:clamp}
\end{figure}

To secure the silicon cantilever, a three-plate clamp system of $\#$8 mirror polished stainless steel was designed to provide a uniform pressure distribution on the base of the cantilever as shown in Figure \ref{fig:clamp}. The mirror polished stainless steel reduces clamping losses due to surface roughness and friction. The outer plate is clamped via a cylindrical bearing to a stainless steel pressure distribution plate that clamps the cantilever to the inner stainless steel plate to provide even pressure across the base of the cantilever. This entire stainless steel assembly is clamped with stainless steel bolts and the final pressure plate that clamps the cantilever is bolted separately to the aluminium block (coloured blue in Figure \ref{fig:clamp}). A torque wrench that is not spring loaded is used to tighten the bolts to $4.5$\,Nm. The absence of recoil from the spring avoids an impulse that might break the cantilever. As the entire clamp assembly is made from stainless steel, the only differential thermal expansion/contraction that occurs is due to the thin silicon cantilever. With this design the cantilever should be clamped more securely when it is cooled below room temperature as silicon has a lower coefficient of thermal expansion than stainless steel.

\begin{figure}
\centering
\includegraphics[width=0.7\linewidth]{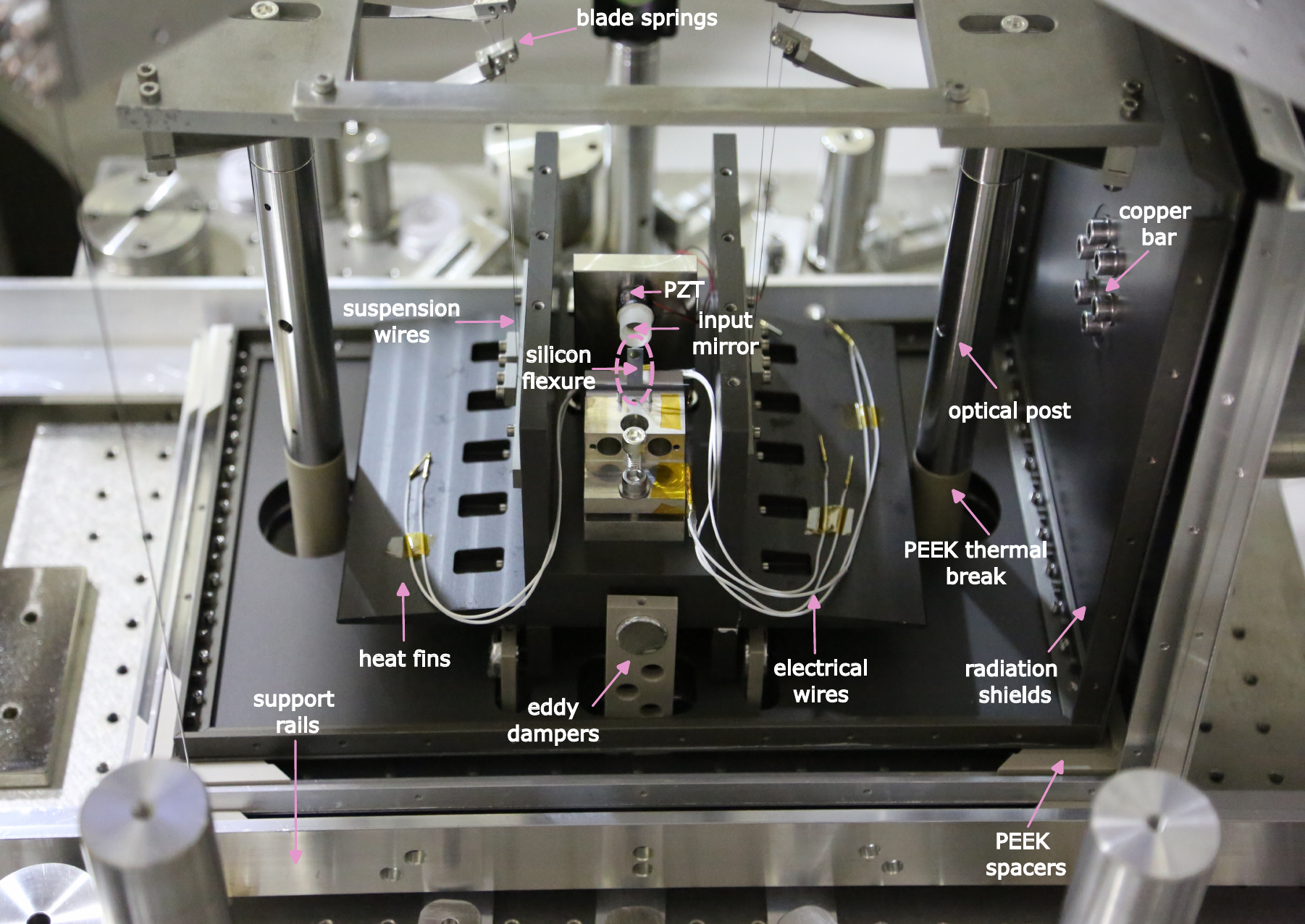}
\caption{Photo showing the test cavity (as illustrated in Figures \ref{fig:testcavity} and \ref{fig:clamp}) and part of the cryogenic infrastructure assembled inside the vacuum tank. The test cavity consists of a curved input mirror and a flat output mirror on a silicon cantilever. The electrical wires leaving the test cavity are used for temperature sensing and control. The cavity is suspended via four stainless steel wires attached to blade springs at the top of the optical posts. The photo shows the three layer radiation shields with the inner one anodised (appearing matte dark grey). PEEK spacers separate the consecutive radiation shields and provide a thermal break for conductive heat flow.}
\label{fig:tctank}
\end{figure}

Stainless steel wires (AISI 316L) with a diameter of 0.3\,mm suspend the TC via four maraging steel blade springs (item (e) in Figure \ref{fig:breadboard}) mounted on an aluminium plate at the top of the optical posts. The optical post assembly consists 1 inch stainless steel posts with PEEK in the middle. At the base, a 1 inch tall stainless steel post is connected via three point contact to a PEEK section passing through the radiation shields with a vented stainless steel M6 grub screw. The reduced contact area, combined with the fact that PEEK is a poor conductor of heat, minimises the heat flow from the room temperature optical breadboard to the TC and acts as a thermal break. 
To damp the motion of the TC, eddy dampers are are attached to the optical posts (item (i) in Figure \ref{fig:breadboard}). Figure \ref{fig:tctank} shows the assembly of the test cavity along with the radiation shields\footnote{In Figure \ref{fig:tctank}, the eddy dampers are of an earlier design that went through the base of the shields. This design proved to be a constraint on the pendulum motion relative to the radiation shields.}.

For readout, the reflected beam from the cavity is sensed on a photodetector and an error signal is derived using the Pound-Drever-Hall (PDH) locking technique \cite{Black2001}. A low-bandwidth feedback loop on the TC PZT (UGF <20\,Hz) is used to account for slow frequency drift. 

\subsection{Temperature sensing and control}
\label{section:tempsensing}

Three cryogenic silicon diodes (S950) and one platinum resistance temperature detector (RTD, 1k$\ohm$) from Cryogenic Control Systems (Cryocon) are used for readout and diagnostics \cite{Cryocon}. These sensors are monitored with the Stanford Research System Cryogenic Temperature Controller (SRS CTC100) that can take 4 four-wire sensor inputs, and has 2 heater outputs capable of delivering 100\,W. 

Two of the silicon diodes are in a bobbin package (UNC 4-40) that are directly screwed into a tapped hole on the ends of the copper bar, i.e., one near the coldhead and the other near the inner shield. 
The third silicon diode is 
in a surface mount package with a footprint of 1\,mm. This surface mount diode was specially used for reading the temperature of the silicon cantilever as indicated in Figure \ref{fig:clamp}. The diode was carefully glued close to the silicon cantilever on the inner stainless steel plate that is used for clamping it. A two-part glass-filled cryogenic epoxy (Stycast 2850FT) was used to bond the diode to the clamp. This epoxy has low thermal expansion, low outgassing and high thermal conductivity, all of which are critical in a vacuum cryogenic environment \cite{Stycast, Stycast2}. 
The final sensor, a 1\,k$\ohm$ platinum RTD, was installed on the far side of the inner radiation shield for diagnostics. It is important to mention that all electrical wires that connect to the TC assembly run along the seismic isolation chain with significant slack thereby reducing mechanical short-circuiting of the suspension system. 

\begin{figure}
\centering
\includegraphics[width=0.5\linewidth]{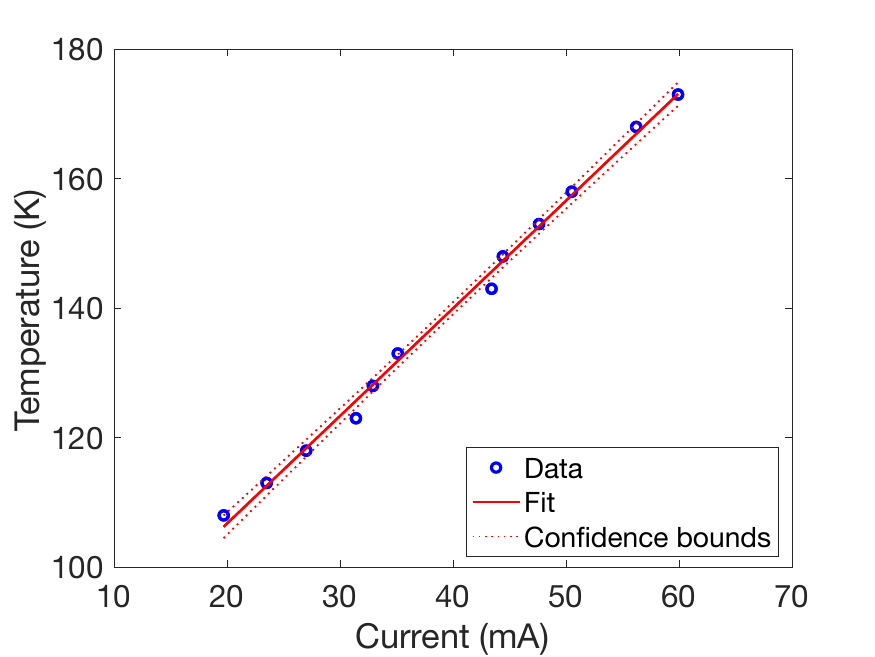}
\caption{Data and linear fit between heater current and the sensor reading the temperature of the silicon cantilever with the 95$\%$ confidence interval also shown.}
\label{fig:tempvsI}
\end{figure}

The outer stainless steel plate of the clamp has a standard 250\,$\ohm$ resistor mounted on it with 
a thin layer of indium foil as shown in Figure \ref{fig:clamp}. This resistor is used to deliver heater power to the cavity for temperature control and stabilisation. 
The total resistance of the resistor and the connecting wire as measured at the temperature controller using a digital multimeter is 263\,$\ohm$. 
 
The resistor and the silicon diode sensor near the cantilever are used in a feedback loop to hold the temperature of the TC at any given set point between 90\,K and 250\,K to $\pm$1\,mK as measured at the cantilever. Where possible, vacuum and cryogenic compatible heat shrink tubing is used for insulation purposes and when the spatial constraints were tight, kapton tape was used to provide the necessary insulation. Figure \ref{fig:tempvsI} shows the variation of the heater current with respect to the set point temperature of the optical TC with the cryostat on. From the graph we can see that the current has a linear dependence with the temperature, with a slope of 1.66\,$\pm$\,0.07\,K/mA.

\begin{figure}{t!}
\centering
\includegraphics[width=0.7\linewidth]{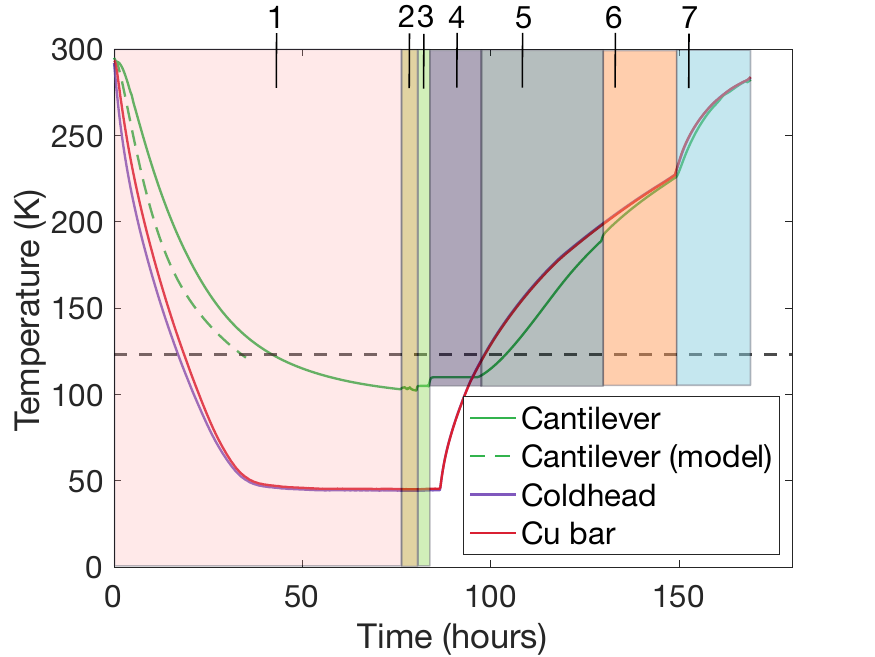}
\caption{First cooldown of the cryogenic infrastructure to test that the facility can go well beyond the 123\,K requirement. Region 1: Cooldown to 123\,K takes around 41\,hours. Region 2: Setting up the feedback loop for temperature control. Region 3: Cavity locked to 105\,K with active feedback (cryostat on). Region 4: Cavity locked to 110\,K with active feedback after the cryostat is turned off. Region 5: Warm up of the cryogenic infrastructure. Region 6: Heater turned on to see the effect on the warm up of the cavity. Region 7: Dry nitrogen gas is pumped into the chamber to reach ambient temperature faster. The green dashed line is the predicted rate of cooling from the cooldown model for the cantilever, taken from Figure \ref{fig:thermalmodel}.}
\label{fig:cooldown2021}
\end{figure} 

\section{Results and discussion}
\label{section:results}

This section presents the results from the cryogenic facility. In particular, we discuss the long-term performance of the facility and the first broadband displacement noise measurements from a crystalline silicon cantilever at 123\,K and compare it against known room temperature measurements. 

\subsection{Performance of the cryogenic facility}
\label{section:performance}

Figure \ref{fig:cooldown2021} shows the results from the first cryogenic run of this facility. The Coldhead and Cu bar curves show the temperature data from the sensors screwed onto the copper bar. One sensor is near the coldhead and the other sensor is near the radiation shield, separated by length of the copper bar. The Coldhead and Cu bar sensors reach a minimum temperature of 44\,K and 45\,K respectively. The green curve is the temperature data from the silicon diode that is installed near the cantilever. The data shows that the cantilever takes $\sim$41 hours to radiatively cool down to 123\,K. This is in good agreement with the cooldown model discussed in section \ref{section:thermalmodel} which predicts a cooldown time of 35 hours for the cantilever (see Figure \ref{fig:cooldown2021}). The discrepancy between the model and data may be attributed to uncertainties in geometric shape factors, conduction at joints and emissivity of the radiation shields. The various shaded regions on the plot highlight cooldown, setting up of feedback loop, and warming up of the infrastructure. It is important to note that when the compressor is turned off after 86 hours, it takes about 11 hours for the coldhead to reach 123\,K. This is because the large copper bar acts as a thermal reservoir keeping the cavity cold for an extended duration. This allows for undertaking low-noise measurements of the silicon cantilever without additional vibration from the compressor and is a key highlight of this cryogenic facility.

\begin{figure}
\centering
\includegraphics[width=0.7\linewidth]{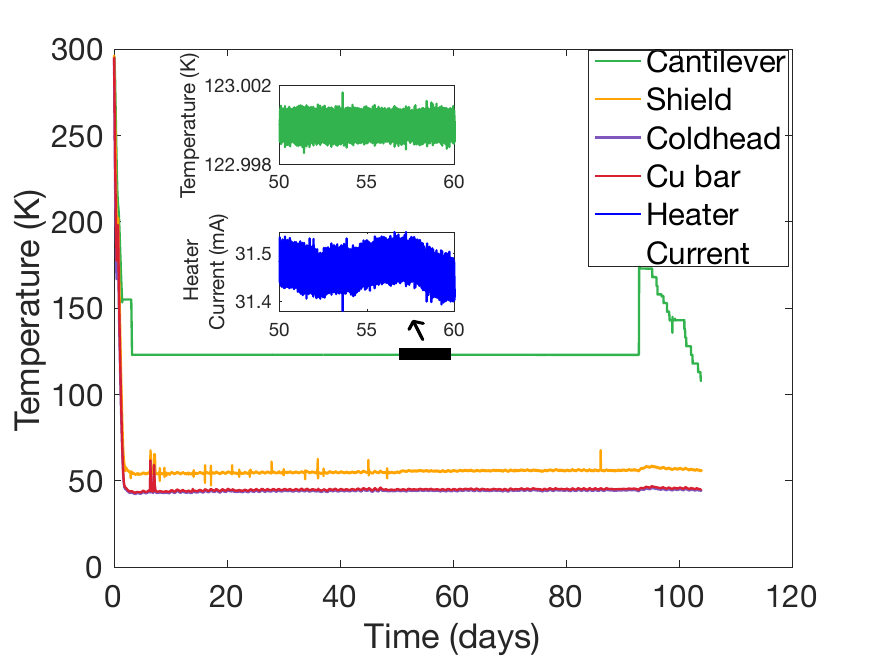}
\caption{Long-term stability of the cryogenic infrastructure over a period of $\sim$100 days. Beyond day 3, the Coldhead and Cu bar sensors mounted on the copper bar settle at $\sim$45\,K whereas the Shield sensor bottoms out at $\sim$55\,K. The inset shows the temperature and heater current stability over a 10-day period. The temperature is stable to $\pm$1\,mK around 123\,K. Beyond day 90, the cavity temperature was changed with active feedback control to allow for ringdown measurements.}
\label{fig:cooldown2023}
\end{figure}

Figure \ref{fig:cooldown2023} shows the long-term performance of the cryogenic infrastructure. The Coldhead and Cu bar sensors remain stable at 44\,K and 45\,K respectively. The two peaks at around day 6 correspond to turning off the cryogenic compressor for low-noise measurements of the silicon cantilever and diagnostics. Temperature data from the shield sensor shows that it reaches a minimum temperature of 55\,K. This shows that the inner radiation shield which is conductively cooled does not lag significantly with respect to the copper bar. The inset plot shows the relative temperature stability of the cantilever to 123\,K\,$\pm$\,1\,mK (reaching the sensor noise floor) and this sensitivity is consistent with an open loop measurement taken independently. This stability allows us to lock to an optical resonance of an optical cavity to make interferometric temperature-dependent displacement measurements between 90\,K and 250\,K. Beyond day 90, the temperature of the cavity was changed to make temperature-dependent measurements from the cantilever.

\subsection{Cryogenic displacement noise measurement}
\label{section:displacement}

\begin{figure}
\centering
\includegraphics[width=0.7\linewidth]{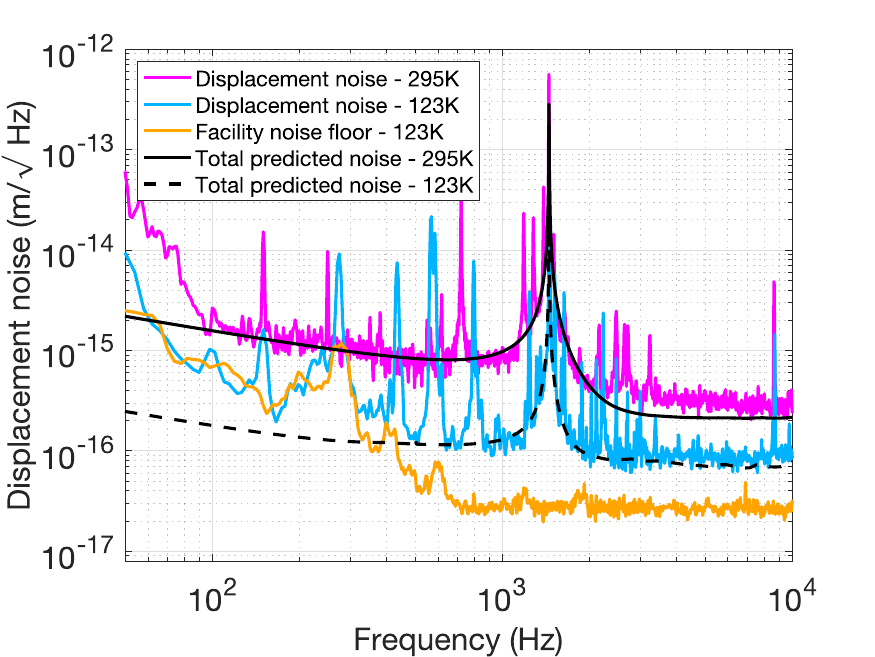}
\caption{Displacement noise spectra from a 500\,$\mu$m silicon cantilever with the fundamental mechanical resonance at 1450\,Hz. The data is recorded at 295\,K and 123\,K and the facility noise floor at 123\,K is shown.}
\label{fig:1450flexure}
\end{figure}

The cryogenic facility was designed for the purpose of measuring broadband displacement noise of a silicon cantilever around the zero crossing of the coefficient of thermal expansion at 123\,K. Figure \ref{fig:1450flexure} shows the displacement noise measured at 295\,K and at 123\,K from such a cantilever. From the figure, we can see that the total displacement noise significantly decreased with cooling, and the measured data approximately agrees with the model for both temperatures. The black lines represents the total predicted noise which is the quadrature sum of all thermal noises \cite{KapasiThesis2024}. The regions above and below the resonance in Figure \ref{fig:1450flexure} show multiple peaks that contaminates the displacement noise spectra. A potential reason for the peaks could be resonant coupling between the cantilever and the clamp, and the cantilever with the heat fins on the TC mount. The graph also highlights that the noise floor of the cryogenic facility has a clearance from the displacement noise at the level of $3 \times 10^{-17} \mathrm{{m}/{\sqrt{Hz}}}$ above 700\,Hz. The facility noise floor shown here is the quadrature sum of the offresonance and the photodetector dark noise. The \lq offresonance\rq\, trace is the optical signal from the cavity when it is sufficiently far away ($\sim$hundreds of optical linewidth) from any cavity modes. For this, the DC offset on the PZT is adjusted such that the cavity sits on the opposite side to where the higher order modes (such as $\mathrm{TEM_{10}}$) occurs. This corresponds to maximum DC optical power on the reflected photodetector, and holds information about cyclic error due to scattering between the laser and the test cavity. In Figure \ref{fig:1450flexure}, below 400\,Hz the displacement noise at 123\,K is affected by cyclic error due to stray light reflections and hence cannot measure thermal noise. 

A decrease in the total noise leads to a higher quality factor ($Q$), and the inverse of $Q$ is a measure of the total loss. The total loss of an object is related to its thermodynamic fluctuations through the Fluctuation Dissipation Theorem \cite{Callen:1952}. The  overall trend in the improvement of loss with cooling is in general agreement with previously measured loss angles in silicon through ringdown measurements \cite{Nawrodt:2013uca,Reid:2004yqx,Nawrodt:2008}. 
Here, the measured total loss values are estimated through ringdown measurements at the fundamental resonant frequency and are summarised in Table \ref{table:1450tempdata}.  Given the high loss values as compared to others, we speculate that the displacement noise is limited by clamping loss. Previous works have reported that fabricating a silicon cantilever with a flexing membrane such that the thickness ratio between the base of the cantilever to the flexing region is $\sim$5:1 can help isolate the thermal driven fluctuations in the bulk material by minimising the dissipation of energy into the clamp structure \cite{ThanhKapasi:2023, Nawrodt:2013uca}. 

\begin{table}[h!]
\centering
\begin{tabular}{| c | c | c | l |} \hline
Temperature (K) & Total loss $\phi_{0}(\omega_0)$\\\hline
295 & $1.96 \times 10^{-4} \pm 3.95 \times 10^{-5}$\\
123 & $1.65 \times 10^{-5} \pm 1.08 \times 10^{-5}$\\   
\hline
\end{tabular}
\caption{Total loss from a 500\,$\mu$m-thick dual side polished silicon cantilever at 295\,K and 123\,K at the fundamental resonant frequency.}
\label{table:1450tempdata}
\end{table}



\section{Conclusion}
\label{section:conclusion}
In this paper we presented the design, commissioning and performance of a low-noise cryogenic facility encompassing an existing seismic isolation chain. This allowed us to measure displacement noise in crystalline silicon at 123\,K where the coefficient of thermal expansion crosses zero. The facility demonstrated radiatively cooling a volume of $\sim36$ litres in 41 hours to 123\,K and has a noise floor at the level of $3 \times 10^{-17}$ m/$\rm{\sqrt{Hz}}$ in the audio-frequency band where current and proposed future ground-based gravitational-wave detectors are most sensitive. More work is needed to measure the thermal noise in crystalline silicon at 123\,K. It would be necessary to design a cantilever which minimises mechanical coupling to the clamp. Based on our previous room temperature results, a cantilever with a membrane etched to nanometre surface roughness is required \cite{ThanhKapasi:2023}. Low-noise resonant photodetectors and finite element analysis of the test cavity mount to identify the mechanical resonances contaminating the spectra would also be required. The displacement data shown here emphasises that using silicon for suspensions in cryogenic gravitational-wave detectors requires careful design for low suspension thermal noise. The improvement in mechanical loss at cryogenic temperatures highlights that silicon is a promising material for future detectors and significant work is needed to lower the total loss. A further application of this cryogenic facility could be to measure the broadband coating noise in promising coating candidates for future ground-based gravitational-wave detectors. Such an experiment will provide valuable knowledge that can help identify coating candidates for implementation in full-scale interferometers. A study like this can also benefit optical cavities requiring an ultra low noise spacer as a frequency reference that are limited by coating thermal noise. 

\ack{Acknowledgments}
This work was supported by the Australian Research Council
(No. DP160100760). The authors would like to thank Professor
G. D. Hammond and Professor S. Rowan for useful discussions. Additionally, we express our gratitude to Dr Muhammad Hassan Iqbal and Mr Deon Hickey for technical assistance.

\end{document}